\documentclass[letterpaper,12pt,notoc]{JHEP3}
\def\pd{\partial}
\def\mc{\mathcal}

\usepackage{graphicx}
\usepackage{amsmath}
\usepackage{amssymb}
\preprint{ \hbox{}\hfill arXiv: 1509.07431}

\title{Non-semisimple gauging of a magical $N=4$ supergravity in three dimensions}
\author{Parinya Karndumri\\
String Theory and Supergravity Group, Department
of Physics, Faculty of Science, Chulalongkorn University, 254 Phayathai Road, Pathumwan, Bangkok 10330, Thailand\\
E-mail: \email{parinya.ka@hotmail.com}}

\abstract{We construct a new $N=4$ non-semisimple gauged
supergravity in three dimensions with $E_{6(2)}/SU(6)\times SU(2)$
scalar manifold and $SO(4)\ltimes \mathbf{T}^6$ gauge group.
Depending on the values of the gauge coupling constants, the theory
admits both the maximally supersymmetric $AdS_3$ vacuum preserving
$SO(4)$ gauge symmetry and half-supersymmetric domain walls with
unbroken $SO(4)$ symmetry. We give all scalar masses at the
supersymmetric $AdS_3$ critical point corresponding to an $N=(4,0)$
superconformal field theory (SCFT) in two dimensions. The scalar
potential also admits two flat directions corresponding to marginal
deformations that preserve full supersymmetry and conformal
symmetry. This $SO(4)\ltimes \mathbf{T}^6$ gauged supergravity is
expected to arise from a dimensional reduction on a three-sphere of
the minimal $N=(1,0)$ supergravity in six dimensions coupled to
three tensor and four vector multiplets.}


\begin{document}
\section{Introduction}
Gauged supergravities in various dimensions play an important role
in many aspects of string/M theory in particular the AdS/CFT
correspondence \cite{maldacena}. Unlike higher dimensional
analogues, gauged supergravity in three dimensions has a much richer
structure due to the duality between vectors and scalars. Since
three-dimensional supergravity fields are topological, the
propagating bosonic degrees of freedom of the matter-coupled
supergravity can be described entirely in terms of scalar fields.
The resulting theory is a supersymmetric non-linear sigma model
coupled to supergravity. All of these theories with different
numbers ($N$) of supersymmetries have been classified in
\cite{dewit1}. For $N>4$, scalar fields must be described by a
symmetric space of the form $G/H$ in which $G$ and $H$ are global
and local symmetries, respectively. The latter is the maximal
compact subgroup of $G$ and takes the form $SO(N)\times H'$ with
$SO(N)$ being the R-symmetry group.
\\
\indent Gauged supergravity can be constructed by introducing gauge
vector fields via Chern-Simons (CS) terms which are topological in
nature. This CS formulation makes the connection to usual Yang-Mills
(YM) gauged supergravities, in which gauge fields appear via YM
kinetic terms, obscure. Furthermore, since conventional dimensional
reductions result in YM gauged supergravities, the embedding of
these CS gauged supergravities in higher dimensions is then a
non-trivial task. However, it has been shown that CS gauged
supergravity with a non-semisimple gauge group $G_0\ltimes
\mathbf{T}^{\textrm{dim}G_0}$ is on-shell equivalent to YM gauged
supergravity with $G_0$ gauge group \cite{csym}. The
$\mathbf{T}^{\textrm{dim}G_0}$ factor is a nilpotent translational
group transforming in the adjoint representation of a semisimple
group $G_0$. Therefore, at least some of these particular gauge
groups might have higher dimensional origins in terms of known
dimensional reductions. Since the embedding in higher dimensions is
necessary for interpreting three-dimensional solutions in
string/M-theory context, CS gauged supergravities with
non-semisimple gauge groups are of particular interest in the
AdS/CFT correspondence.
\\
\indent Until now, a number of works considering CS gauged
supergravity with non-semisimple gauge groups and their application
in the AdS/CFT correspondence have appeared
\cite{nicolai3,gkn,henning_AdS3_S3,henning_hohm,N4_gauging,DW3D,1_2_DW,LargeN44,N5_6}.
All of these gauged supergravities are expected to arise from
dimensional reductions of higher dimensional theories, and explicit
reduction ansatze for obtaining three-dimensional gauged
supergravities from a three-sphere reduction, both the full $S^3$
reduction and $SU(2)$ group manifold reduction, have been
constructed in \cite{PopeSU2,PopeSU22,KK,Henning_S3_to_3D}.
Furthermore, a number of $N=2$ gauged supergravities with abelian
gauge groups have also been obtained from wrapped branes in type IIB
and M-theory \cite{henning_ADSCMT,EOC_PKD3,EOC_PKM5}.
\\
\indent In this paper, we construct new $N=4$ gauged supergravity
with $E_{6(2)}/SU(6)\times SU(2)$ scalar manifold and $SO(4)\ltimes
\mathbf{T}^6$ gauge group. According to the result of \cite{csym},
the resulting gauged supergravity is equivalent to $SO(4)$ YM gauged
supergravity. A number of possible semisimple gaugings of this
theory have already been classified in \cite{N4_gauging}. In the
present work, we will study possible non-semisimple gauge groups and
also classify supersymmetric vacua of the gauged supergravity both
maximally supersymmetric and half-supersymmetric with the full
$SO(4)$ symmetry unbroken. These would be useful in AdS$_3$/CFT$_2$
correspondence.
\\
\indent The theory considered here is one of the magical
supergravities in three dimensions, see
\cite{Magic_SUGRA1,Magic_SUGRA2,Romans_6D_magic} for higher
dimensional magical supergravities. In \cite{N4_gauging}, another
$SO(4)\ltimes \mathbf{T}^6$ gauged magical supergravity with
$F_{4(4)}/USp(6)\times SU(2)$ scalar manifold has been studied. This
theory is expected to arise from an $S^3$ reduction of $N=(1,0)$
supergravity in six dimensions coupled to two vector and two tensor
multiplets. In the present case, we expect the new $SO(4)\ltimes
\mathbf{T}^6$ gauged theory with $E_{6(2)}/SU(6)\times SU(2)$ scalar
manifold to arise from an $S^3$ reduction of this $N=(1,0)$
supergravity coupled to four vector and three tensor multiplets.
Both of these six-dimensional supergravities are also magical
supergravities, and the possible gaugings have been systematically
considered in \cite{Henning_6D_magic} using the embedding tensor
formalism.
\\
\indent The paper is organized as follow. In section \ref{N4theory},
we review $N=4$ three-dimensional gauged supergravity with symmetric
scalar target spaces. We mainly focus on a specific case of
exceptional coset $E_{6(2)}/SU(6)\times SU(2)$. In section
\ref{vacua}, the embedding tensor of $SO(4)\ltimes \mathbf{T}^6$ is
given. We then study the resulting scalar potential for the $SO(4)$
singlet scalars and identify some of their possible supersymmetric
vacua. We end the paper by giving some conclusions and comments on a
possible higher dimensional origin of the gauged supergravity
constructed here in section \ref{conclusions}. Two appendices with
some useful formulae are also included.
\section{$N=4$ gauged supergravity in three
dimensions with $E_{6(2)}/SU(6)\times SU(2)$ scalar
manifold}\label{N4theory} Three dimensional matter-coupled
supergravity is given by a nonlinear sigma model coupled to
supergravity. Scalar fields will be denoted by $\phi^i$,
$i=1,\ldots, d$ with $d$ being the dimension of the scalar target
space. Throughout this paper, we will work in the $SO(N)$ covariant
formulation of \cite{dewit}.
\\
\indent Coupling the non-linear sigma model to $N$ extended
supergravity requires $N-1$ almost complex structures, denoted by
$f^{Pi}_{\phantom{sdf}j}$, $P=2,\ldots N$, on the target space of
the sigma model. The R-symmetry in three dimensions is given by
$SO(N)$ under which scalars transform in a spinor representation.
The tensor $f^{IJ}$, $I,J=1,\ldots N$, generating $SO(N)$ R-symmetry
in this spinor representation, can be constructed from $f^P$ via
\begin{equation}
f^{1P}=-f^{P1}=f^P,\qquad f^{PQ}=f^{[P}f^{Q]}\, .\label{fIJ}
\end{equation}
The $f^{IJ}_{ij}$ have a symmetry property
$f^{IJ}_{ij}=-f^{JI}_{ij}=-f^{IJ}_{ji}$.
\\
\indent For $N=4$ theory, the scalar target space must be a
quaternionic manifold. Special to $N=4$ supersymmetry, there exists
a tensor $J=\frac{1}{6}\epsilon_{PQR}f^Pf^Qf^R$ that commutes with
the almost complex structures and is covariantly constant. This
implies the product structure of the target space. Therefore, a
general $N=4$ matter-coupled supergravity admits a scalar manifold
of the form $\mc{M}=\mc{M}_-\times \mc{M}_+$ with a total dimension
$d=d_++d_-$. The full $SO(4)\sim SO(3)_-\times SO(3)_+$ R-symmetry
will split into each factor of the full target space. In this paper,
we are however only interested in the so-called degenerate case with
only one factor of $\mc{M}$ present. For definiteness, we will
consider a non-vanishing $\mc{M}_+$ by setting $d_-=0$. Furthermore,
we will restrict ourselves to only a symmetric target space of the
form $G/H$ although this is not in general a requirement from $N=4$
supersymmetry.
\\
\indent The symmetric target space of the form $G/H$ has a global
symmetry $G$ and a local symmetry $H$ given by its maximal compact
subgroup. In general, the local symmetry $H$ contains the $SO(N)$
R-symmetry and takes the form $H=SO(N)\times H'$. But, for the $N=4$
theory, we have the compact group $H_\pm=SO(3)_\pm\times
H'_{\pm}\sim SU(2)_\pm\times H'_{\pm}$ for each subspace
$\mc{M}_\pm$. In the present case, the scalar target space is given
by $E_{6(2)}/SU(6)\times SU(2)$ of dimension $40$.
\\
\indent We now decompose the $E_{6(2)}$ generators
$t^{\mc{M}}=(T^{IJ}_+,X^{\alpha},Y^A)$ into those of the compact
$SO(3)_+ \times SU(6)$ and the non-compact generators $Y^A$,
$A=1,\ldots, 40$. The $SO(3)_+$ generators denoted by $T^{IJ}_+$ are
given by the self-dual part of the full $SO(4)$ R-symmetry
generators $T^{IJ}$:
\begin{equation}
T^{IJ}_{+}=T^{IJ}+\frac{1}{2}\epsilon_{IJKL}T^{KL}\, .
\end{equation}
$X^\alpha$, $\alpha=1,\ldots, 35$ are $SU(6)$ generators.
\\
\indent The $E_{6(2)}/SU(6)\times SU(2)$ manifold can be described
by the coset representative $L$ transforming under $E_{6(2)}$ and
$SU(6)\times SU(2)$ by left and right multiplications, respectively.
In particular, $L$ can be used to find the $SU(6)\times SU(2)$
composite connections, $Q_{+i}^{IJ}$ and $Q^\alpha_i$, and the
vielbein on $E_{6(2)}/SU(6)\times SU(2)$, $e^A_i$, by the relation
\begin{equation}
L^{-1} \partial_i L= \frac{1}{2}Q^{IJ}_{+i} T^{IJ}_++Q^\alpha_i
X^{\alpha}+e^A_i Y^A\, .\label{cosetFormula1}
\end{equation}
The metric on the target space $g_{ij}$ can be computed from the
vielbein $e^A_i$ by the usual relation
$g_{ij}=e^A_ie^B_j\delta_{AB}$. Here, indices $A,B=1,\ldots 40$ can
be considered as ``flat'' target space indices.
\\
\indent Gaugings are implemented by a symmetric and gauge invariant
tensor called the embedding tensor $\Theta_{\mathcal{MN}}$
\cite{nicolai1,N8}. A viable gauging consistent with supersymmetry
is characterized by the embedding tensor that satisfies two
consistency conditions. The first condition, called the quadratic
constraint, requires that the gauge group is a proper subgroup of
the global symmetry $G$. This constraint is explicitly given by
\begin{equation}
\Theta_{\mathcal{PL}}f^{\mathcal{KL}}_{\phantom{ads}\mathcal{(M}}\Theta_{\mathcal{N)K}}=0\label{theta_quadratic}
\end{equation}
where $f^{\mathcal{KL}}_{\phantom{ads}\mathcal{M}}$ are the
$G$-structure constants. Furthermore, supersymmetry requires that
the T-tensor defined by the image of the embedding tensor under a
map $\mc{V}$
\begin{equation}
T_{\mathcal{AB}}=\mathcal{V}^{\mc{M}}_{\phantom{as}\mc{A}}\Theta_{\mc{MN}}\mathcal{V}^{\mc{N}}_{\phantom{as}\mc{B}}\label{T_tensor_def}
\end{equation}
satisfies the constraint
\begin{equation}
T^{IJ,KL}=T^{[IJ,KL]}-\frac{4}{N-2}\delta^{[I[K}T^{L]M,MJ]}-\frac{2}{(N-1)(N-2)}\delta^{I[K}\delta^{L]J}T^{MN,MN}\,
.\label{Tconstraint}
\end{equation}
The T-tensor transforms under the local $H$ symmetry with the index
$\mathcal{A}=\{IJ,\alpha,A\}$. The above constraint implies that the
$\boxplus$ representation of $SO(N)$ is absent from the $T^{IJ,KL}$
component of the T-tensor. Therefore, we can equivalently write this
constraint as
\begin{equation}
\mathbb{P}_\boxplus T^{IJ,KL}=0\, .\label{Tconstraint1}
\end{equation}
\indent In the case of symmetric target spaces, the condition
\eqref{Tconstraint1} can be expressed as a consistency condition on
the embedding tensor which lives in a symmetric product of the
adjoint representation of $G$. It turns out that under the
decomposition of this product into irreducible representations of
$G$, there is a unique representation of $G$, called $R_0$, which
gives rise to the $SO(N)$ representation $\boxplus$ under the
branching $G\rightarrow SO(N)$. Therefore, in this case, the
constraint \eqref{Tconstraint1} can be written in a $G$-covariant
way by
\begin{equation}
\mathbb{P}_{R_0}\Theta_{\mc{MN}}=0\, .\label{Theta_constraint}
\end{equation}
In the case of $G=E_{6(2)}$, the representation $R_0$ is given by
$\textbf{2430}$ \cite{dewit}. In addition, for symmetric spaces in
the form of a coset space, the map $\mc{V}$ will be an isomorphism,
and its components are given by the relation
\begin{equation}
L^{-1}t^\mathcal{M}L=\frac{1}{2}\mathcal{V}^{\mathcal{M}}_{\phantom{as}IJ}T^{IJ}_++\mathcal{V}^\mathcal{M}_{\phantom{as}\alpha}X^\alpha+
\mathcal{V}^\mathcal{M}_{\phantom{as}A}Y^A\, .\label{cosetFormula}
\end{equation}
\indent Various components of the T-tensor can be straightforwardly
computed from the embedding tensor by using the definition
\eqref{T_tensor_def} and the map $\mc{V}$ from \eqref{cosetFormula}.
Combinations of these T-tensor components are used to construct the
$A_1$, $A_2$ and $A_3$ tensors which will appear as fermion
mass-like terms and the scalar potential in the gauged Lagrangian.
They are defined by
\begin{eqnarray}
A_1^{IJ}&=&-\frac{4}{N-2}T^{IM,JM}+\frac{2}{(N-1)(N-2)}\delta^{IJ}T^{MN,MN},\\
A_{2j}^{IJ}&=&\frac{2}{N}T^{IJ}_{\phantom{as}j}+\frac{4}{N(N-2)}f^{M(I
m}_{\phantom{as}j}T^{J)M}_{\phantom{JM}m}+\frac{2}{N(N-1)(N-2)}\delta^{IJ}f^{KL\phantom{a}m}_{\phantom{KL}j}T^{KL}_{\phantom{KL}m}, \phantom{asad}\\
A^{IJ}_{3ij}&=&\frac{1}{N^2}\left[-2D_{(i}D_{j)}A^{IJ}_1+g_{ij}A^{IJ}_1+A_1^{K[I}f^{J]K}_{ij}
+2T_{ij}\delta^{IJ}-4D_{[i}T^{IJ}_{j]}
-2T_{k[i}f^{IJk}_{\phantom{IJk}j]}\right]\nonumber \\
& &
\end{eqnarray}
where $D_i$ is the covariant derivative with respect to $\phi_i$. In
terms of these tensors, the scalar potential can be written as
\begin{equation}
V=-\frac{4}{N}\left(A_1^{IJ}A_1^{IJ}-\frac{1}{2}Ng^{ij}A_{2i}^{IJ}A_{2j}^{IJ}\right).
\end{equation}
\\
\indent As a final ingredient, we give the supersymmetry
transformations of the gravitini $\psi^I_\mu$ and the spinor fields
$\chi^{iI}$, involving only bosonic fields,
\begin{eqnarray}
\delta\psi^I_\mu
&=&\mathcal{D}_\mu\epsilon^I+gA_1^{IJ}\gamma_\mu\epsilon^J,\label{d_psi}\\
\delta\chi^{iI}&=&
\frac{1}{2}(\delta^{IJ}\mathbf{1}-f^{IJ})^i_{\phantom{a}j}{\mathcal{D}{\!\!\!\!/}}\phi^j\epsilon^J
-gNA_2^{JIi}\epsilon^J\label{d_chi}
\end{eqnarray}
where the covariant derivative of $\epsilon^I$ is given by
\begin{equation}
\mc{D}_\mu
\epsilon^I=\pd_\mu\epsilon^I+\frac{1}{4}\omega_\mu^{ab}\gamma_{ab}\epsilon^I+\pd_\mu\phi^i
Q_i^{IJ}\epsilon^J+g\Theta_{\mc{MN}}A^\mc{M}_\mu
\mc{V}^{\mc{N}IJ}\epsilon^J\, .
\end{equation}
The covariant derivative on scalars $\phi^i$ is defined by
\begin{equation}
\mc{D}_\mu\phi^i=\pd_\mu \phi^i+\Theta_{\mc{M}\mc{N}}A^{\mc{M}}_\mu
X^{\mc{N}i}
\end{equation}
with $X^{\mc{N}i}$ being the Killing vectors associated to the
isometries of $G/H$.
\\
\indent Note that there are only $d$ physical $\chi^i$ fields in
consistent with $d$ scalar fields $\phi^i$ as required by
supersymmetry. In order to work with the $SO(N)$ covariant
formulation of \cite{dewit}, in which the explicit dependence on
$f^{P}$ is absent, the $\chi^{i}$ fields have been written in terms
of the constrained fields $\chi^{iI}$ satisfying
\begin{equation}
\chi^{iI}=\frac{1}{N}(\delta^{IJ}\delta^i_j-f^{IJi}_{\phantom{IJi}j})\chi^{jJ}\,
.
\end{equation}
We finally note here that, for maximally symmetric vacua, the
unbroken supersymmetry corresponds to Killing spinors $\epsilon^I$
satisfying the relation
\begin{equation}
A_1^{IK}A_1^{KJ}\epsilon^J=-\frac{V_0}{4}\epsilon^I\,.\label{SUSY_condition}
\end{equation}
This relation can be derived by solving the BPS conditions
$\delta\psi^I_\mu=0$ and $\delta\chi^{iI}=0$ at constant scalars,
see the relevant detail in \cite{dewit}.

\section{$N=4$, $SO(4)\ltimes \mathbf{T}^6$ gauged supergravity and some supersymmetric vacua}\label{vacua}
We first give an explicit construction of the $E_{6(2)}/SU(6)\times
SU(2)$ coset space. Generators of the compact $H=SU(6)\times SU(2)$
subgroup and the $40$ non-compact generators are given in appendix
\ref{E6_generator}.
\\
\indent We first describe the embedding of $SO(4)\ltimes
\mathbf{T}^6$ gauge group in the global symmetry group $E_{6(2)}$.
In order to do this, we will decompose $E_{6(2)}$ into its maximal
subgroup $SO(6,4)\times U(1)$ generated by
$\tilde{X}^{\hat{i}\hat{j}}$ and $\hat{X}$ given explicitly in
appendix \ref{E6_generator}. The full gauge group $SO(4)\ltimes
\mathbf{T}^6$ can be embedded in $SO(6,4)$ as follow.
\\
\indent The semisimple part $SO(4)$ is given by a diagonal subgroup
of $SO(4)\times SO(4)\subset SO(6)\times SO(4)$ which is in turn the
maximal compact subgroup of $SO(6,4)$. This $SO(4)$ is accordingly
generated by
\begin{equation}
J^{ab}=\tilde{X}^{ab}+\tilde{X}^{a+6,b+6},\qquad a,b=1,\ldots, 4\, .
\end{equation}
The other combination $\tilde{X}^{ab}-\tilde{X}^{a+6,b+6}$ together
with a suitable set of non-compact generators will give rise to the
translational generators $\mathbf{T}^6$ transforming as an adjoint
representation of the above mentioned $SO(4)$. All of the
$\mathbf{T}^6$ generators also commute with each other.
\\
\indent In order to identify the appropriate non-compact generators
constituting $\mathbf{T}^6$, we decompose the $Y^A$ generators under
the $SO(4)$ part of the gauge group. Under $SU(6)\times SU(2)$, the
$40$ generators $Y^A$ transform as $(\mathbf{20},\mathbf{2})$. Under
$SU(6)\times SU(2)\rightarrow SU(4)\times SU(2)\times U(1)\times
SU(2)$, we find
\begin{equation}
(\mathbf{20},\mathbf{2})\rightarrow
(\mathbf{4}_3,\mathbf{1}_{3},\mathbf{2})+(\bar{\mathbf{4}}_{-3},\mathbf{1}_{-3},\mathbf{2})
+(\mathbf{6}_0,\mathbf{2}_{0},\mathbf{2}).
\end{equation}
From now on, we will neglect all the $U(1)$ charges since they will
not play any important role in the following analysis. We now
decompose $SU(4)\sim SO(6)\rightarrow SO(4)\times SO(2)$ by the
embedding $\mathbf{6}\rightarrow \mathbf{4}+\mathbf{1}+\mathbf{1}$.
With the $SO(2)\sim U(1)$ charges neglected, further decompositions
to $SO(4)\times SO(4)$ and finally to $SO(4)_{\textrm{diag}}$
respectively give
\begin{eqnarray}
SO(4)\times SO(4)&:& 2\times
(\mathbf{1},\mathbf{2};\mathbf{1},\mathbf{2})+2\times
(\mathbf{2},\mathbf{1};\mathbf{1},\mathbf{2})+(\mathbf{2},\mathbf{2};\mathbf{2},\mathbf{2})
+(\mathbf{1},\mathbf{1};\mathbf{2},\mathbf{2})\nonumber \\
SO(4)_{\textrm{diag}}&:&3\times (\mathbf{1},\mathbf{1})+2\times (\mathbf{1},\mathbf{3})+4\times (\mathbf{2},\mathbf{2})+(\mathbf{3},\mathbf{3})\nonumber \\
& &+(\mathbf{1},\mathbf{3})+(\mathbf{3},\mathbf{1})\label{SO4_rep}
\end{eqnarray}
where we have denoted the $SO(4)\sim SU(2)\times SU(2)$
representations by $(\mathbf{2j_1+1},\mathbf{2j_2+1})$ with
$j_1,j_2$ corresponding to the spins of the two $SU(2)$'s.
\\
\indent The last two representations in \eqref{SO4_rep} are the
adjoint representation of $SO(4)_{\textrm{diag}}$. These will be
part of the gauge generators $\mathbf{T}^6$. Explicitly, we find
that these generators are given by
\begin{equation}
t^{ab}=\tilde{X}^{ab}-\tilde{X}^{a+6,b+6}+\tilde{X}^{a,b+6}+\tilde{X}^{a+6,b},\qquad
a,b=1,\ldots, 4\, .
\end{equation}
Note that $\tilde{X}^{a,b+6}$ and $\tilde{X}^{a+6,b}$ are
non-compact generators of $E_{6(2)}$. It can be verified that
$(J^{ab}, t^{ab})$ generators satisfy the $SO(4)\ltimes\mathbf{T}^6$
algebra
\begin{equation}
\left[J^{ab},J^{cd}\right]=-4\delta^{[a[c}J^{d]b]},\qquad
\left[J^{ab},t^{cd}\right]=-4\delta^{[a[c}t^{d]b]},\qquad
\left[t^{ab},t^{cd}\right]=0\, .
\end{equation}
\indent The non-vanishing components of the embedding tensor
corresponding to the full gauge group are denoted by
$\Theta_{\mathbf{a}\mathbf{b}}$ and $\Theta_{\mathbf{b}\mathbf{b}}$
with $\mathbf{a}$ and $\mathbf{b}$ associated to the $J^{ab}$ and
$t^{ab}$ parts, respectively \cite{csym}. It turns out that the
embedding tensor satisfying the linear and quadratic constraints
given in \eqref{theta_quadratic} and \eqref{Theta_constraint} is
given by
\begin{equation}
\Theta=g_1\Theta_{\mathbf{a}\mathbf{b}}+g_2\Theta_{\mathbf{b}\mathbf{b}}
\end{equation}
where both $\Theta^{\mathbf{a}\mathbf{b}}_{ab,cd}$ and
$\Theta^{\mathbf{b}\mathbf{b}}_{ab,cd}$ are given by
$\epsilon_{abcd}$. This is much similar to the $N=8$ and $N=4$
theories studied in \cite{henning_AdS3_S3} and \cite{N4_gauging}. It
should be noted that supersymmetry allows for two independent
coupling constants.

\subsection{Maximally supersymmetric vacua with $SO(4)$ symmetry}
We now consider some vacua of the $N=4$ gauged supergravity
constructed previously. From equation \eqref{SO4_rep}, we see that
there are three scalars which are singlets under $SO(4)$. These
singlets correspond to the following non-compact generators
\begin{eqnarray}
\hat{Y}_1&=&Y_1-Y_4+Y_5-Y_7-Y_9-Y_{12}+Y_{13}+Y_{15},\nonumber \\
\hat{Y}_2&=&Y_2+Y_3-Y_6+Y_8+Y_{10}-Y_{11}-Y_{14}-Y_{16},\nonumber \\
\hat{Y}_3&=&Y_{17}+Y_{19}+Y_{24}+Y_{26}+Y_{29}-Y_{31}+Y_{36}-Y_{38}\,
.
\end{eqnarray}
\indent The coset representative $L$ can be parametrized by
\begin{equation}
L=e^{\Phi_1\hat{Y}_1}e^{\Phi_2\hat{Y}_2}e^{\Phi_3\hat{Y}_3}\, .
\end{equation}
By using the formulae in section \ref{N4theory} and appendix
\ref{detail}, we obtain the scalar potential for
$(\Phi_1,\Phi_2,\Phi_3)$ given by
\begin{eqnarray}
V&=&32g_1\left[4\cosh(2\Phi_1)\cosh(2\Phi_2)\cosh(2\Phi_3)-4\sinh(2\Phi_3)\right]^2\times\nonumber \\
&
&\left[4g_2\cosh(2\Phi_1)\cosh(2\Phi_2)\cosh(2\Phi_3)-4g_2\sinh(2\Phi_3)-6g_1\right].\label{singlet_potential}
\end{eqnarray}
\indent From this potential, it can be readily verified that the
$SO(4)\ltimes \mathbf{T}^6$ gauged supergravity admits a maximally
supersymmetric $AdS_3$ critical point. By shifting the values of
scalar fields at the vacuum, we can bring the critical point to
$L=\mathbf{I}_{27\times 27}$ at which $\Phi_1=\Phi_2=\Phi_3=0$. This
can also be achieved by setting $g_2=g_1=g$. The cosmological
constant at the critical point is given by $V_0=-1024g^2$. In our
convention, the $AdS_3$ radius is given by
\begin{equation}
L=\sqrt{-\frac{1}{V_0}}=\frac{1}{32g}
\end{equation}
where we have taken $g>0$ for definiteness.
\\
\indent It can be checked by using the supersymmetry transformations
\eqref{d_psi} and \eqref{d_chi} or the relation
\eqref{SUSY_condition} that this critical point preserves the full
eight supercharges corresponding to $N=(4,0)$ superconformal
symmetry in the dual two-dimensional SCFT. All of the scalar masses
at this critical point are given in table \ref{table1}. In the
table, we have also given the dimensions of the dual operators in
the dual SCFT according to the relation $m^2L^2=\Delta(\Delta-2)$. All of the masses agree with the behavior of the corresponding scalar fields near the $AdS_3$ critical points.
\begin{table}[h]
\centering
\begin{tabular}{|c|c|c|}
  \hline
  $SO(4)$ representations & $m^2L^2\phantom{\frac{1}{2}}$ & $\Delta$  \\ \hline
  $(\mathbf{1},\mathbf{1})$ & $0_{(\times 2)}$, $3$ & $2,3$  \\
  $(\mathbf{1},\mathbf{3})+(\mathbf{3},\mathbf{1})$ & $0_{(\times 6)}$  & $2$ \\
  $(\mathbf{1},\mathbf{3})$ & $-\frac{8}{9}_{(\times 6)}$  & $\frac{2}{3}$, $\frac{4}{3}$ \\
  $(\mathbf{2},\mathbf{2})$ & $-\frac{3}{4}_{(\times 16)}$  & $\frac{1}{2}$, $\frac{3}{2}$ \\
  $(\mathbf{3},\mathbf{3})$ & $-1_{(\times 9)}$  & $1$ \\
  \hline
\end{tabular}
\caption{Scalar masses at the $N=4$ supersymmetric $AdS_3$ critical
point and the corresponding dimensions of the dual
operators}\label{table1}
\end{table}
\\
\indent From the table, we see the presence of six massless scalars
in the adjoint representations of $SO(4)$,
$(\mathbf{1},\mathbf{3})+(\mathbf{3},\mathbf{1})$. These are
Goldstone bosons corresponding to the symmetry breaking
$SO(4)\ltimes \mathbf{T}^6\rightarrow SO(4)$ at the vacuum.
Furthermore, there are additional massless scalars which are
singlets under $SO(4)$. These are expected to correspond to marginal
deformations in the dual SCFT. The deformations preserve all
supersymmetry as well as the full $SO(4)$ symmetry. These
deformations can be given explicitly by the relation
\begin{equation}
\cosh(2\Phi_1)\cosh(2\Phi_2)\cosh(2\Phi_3)=1+\sinh(2\Phi_3).
\end{equation}
When $\Phi_1=\Phi_2=0$, there is only one solution $\Phi_3=0$.
Non-vanishing values of $\Phi_1$ and $\Phi_2$ give rise to the same
value of $V_0$, unbroken $SO(4)$ symmetry and the same number of
supersymmetry. Therefore, $\Phi_1$ and $\Phi_2$ correspond to flat
directions of the potential.
\\
\indent There is another class of vacua given by the relation
\begin{equation}
\cosh(2\Phi_1)\cosh(2\Phi_2)=\tanh(2\Phi_3).
\end{equation}
This gives supersymmetric Minkowski vacua in three dimensions with
$V_0=0$.

\subsection{Half-supersymmetric domain walls}
We now move to half-supersymmetric vacuum solutions. To find these
solutions, we set up the corresponding BPS equations from the
supersymmetry transformations \eqref{d_psi} and \eqref{d_chi}. The
three-dimensional metric is taken to be the standard domain wall
ansatz
\begin{equation}
ds^2=e^{A(r)}dx_{1,1}^2+dr^2\, .\label{metric_ansatz}
\end{equation}
With the projection $\gamma_r\epsilon^I=-\epsilon^I$ corresponding
to $N=(4,0)$ supersymmetry in two dimensions, equations
$\delta\chi^{iI}=0$ and $\delta\psi^I_\mu=0$ for $\mu=0,1$ give
\begin{eqnarray}
\Phi_1'&=&\frac{16\sinh(2\Phi_1)}{\cosh(2\Phi_2)\cosh(2\Phi_3)}\left[g_1-g_2\cosh(2\Phi_1)\cosh(2\Phi_2)
\cosh(2\Phi_3)\right.\nonumber\\ & & \left.+g_2\sinh(2\Phi_3)\right],\label{Eq1}\\
\Phi_2'&=&\frac{16\sinh(2\Phi_2)\cosh(2\Phi_1)}{\cosh(2\Phi_3)}\left[g_1-g_2\cosh(2\Phi_1)\cosh(2\Phi_2)
\cosh(2\Phi_3)\right.\nonumber \\ & & \left.+g_2\sinh(2\Phi_3)\right],\label{Eq2}\\
\Phi_3'&=&16\left[g_1-g_2\cosh(2\Phi_1)\cosh(2\Phi_2)
\cosh(2\Phi_3)+g_2\sinh(2\Phi_3)\right]\times\nonumber \\
& &\left[\cosh(2\Phi_1)\cosh(2\Phi_2)
\sinh(2\Phi_3)-\cosh(2\Phi_3)\right],\label{Eq3}\\
A'&=&32\left[2g_1-g_2\cosh(2\Phi_1)\cosh(2\Phi_2)\cosh(2\Phi_3)+g_2\sinh(2\Phi_3)\right]\times\nonumber \\
&
&\left[\cosh(2\Phi_1)\cosh(2\Phi_2)\cosh(2\Phi_3)-\sinh(2\Phi_3)\right]\label{Eq4}
\end{eqnarray}
where $'$ denotes the $r$-derivative. All of these equations satisfy the second-order field equations. Some details of this verification is given in appendix \ref{field_eq}.
\\
\indent We first consider the case $\Phi_1=\Phi_2=0$. It can be
readily seen that the first two equations are identically satisfied.
We are left with two equations
\begin{eqnarray}
\Phi_3'&=&16e^{-4\Phi_3}(g_2-g_1e^{2\Phi_3}),\label{eq1}\\
A'&=&32e^{-4\Phi_3}(2g_1e^{2\Phi_3}-g_2).\label{eq2}
\end{eqnarray}
For $g_2\neq 0$, the $\Phi_3'$ equation has a critical point at
$\Phi_3=\frac{1}{2}\ln\left[\frac{g_2}{g_1}\right]$ while the $A'$
equation gives $A=\frac{32g_1^2}{g_2}r+C$. This is the maximally
supersymmetric $AdS_3$ critical point identified previously.
\\
\indent Equations \eqref{eq1} and \eqref{eq2} can be solved
explicitly with the corresponding solution
\begin{eqnarray}
A&=&-2\Phi_3-\ln(e^{2\Phi_3}g_1-g_2)+C_1,\nonumber \\
32g_1^2r&=&-g_1e^{2\Phi_3}-g_2\ln (e^{2\Phi_3}g_1-g_2)+C_2\, .
\end{eqnarray}
It should be noted that the integration constants $C_2$ and $C_1$
can be removed by shifting the coordinate $r$ and rescaling the
$x^0$ and $x^1$ coordinates, respectively. The solution interpolates
between $N=4$ $AdS_3$ critical point at $r\rightarrow \infty$ to a
half-supersymmetric domain wall at a finite value of $r$. At large
$|\Phi_3|$, we find
\begin{equation}
\Phi_3\sim \frac{1}{2}\ln (C-32g_1r),\qquad A\sim -\ln(C-32g_1r)
\end{equation}
with $C$ being a constant. We have set $g_2=g_1$ for simplicity. The
metric at $r\sim \frac{C}{32g_1}$ becomes
\begin{equation}
ds^2=(C-32g_1r)^{-2}dx^2_{1,1}+dr^2
\end{equation}
which takes the form of a domain wall. However, the scalar potential
becomes unbounded when $\Phi_3\rightarrow -\infty$. The singularity
of the above metric is then unphysical by the criterion of
\cite{Gubser_singularity}. Therefore, the holographic interpretation
of the solution as an RG flow between an $N=(4,0)$ SCFT and an
$N=(4,0)$ non-conformal field theory in two dimensions cannot be
given at least in the three-dimensional framework. It would be
interesting to further investigate the singularity in the context of
higher dimensions in which this solution is embedded. Note also that
since the operator dual to $\Phi_3$ is irrelevant, we would expect
the $N=(4,0)$ SCFT to appear in the IR.
\\
\indent There is another simple exact solution to equations
\eqref{Eq1}, \eqref{Eq2}, \eqref{Eq3} and \eqref{Eq4} namely
\begin{equation}
\Phi_3=0,\qquad g_2\cosh(2\Phi_1)\cosh(2\Phi_2)=g_1,\qquad
A=\frac{32g_1^2}{g_2}r+C\, .
\end{equation}
This solution corresponds to a marginal deformation of the
supersymmetric $AdS_3$ critical point.
\\
\indent We now move to another type of domain walls which is a
half-supersymmetric vacuum of the theory without any limit with
enhanced supersymmetry. Recall that supersymmetry allows for two
independent gauge couplings $g_1$ and $g_2$, by setting $g_2=0$, we
also have a consistent gauged supergravity. In this case, the
resulting gauged supergravity possesses a half-supersymmetric domain
wall vacuum. We will present a simple solution of this type. With
$\Phi_1=\Phi_2=g_2=0$, the BPS equations become
\begin{eqnarray}
\Phi_3'&=&-16g_1e^{-2\Phi_3},\nonumber\\
A'&=&64g_1e^{-2\Phi_3}
\end{eqnarray}
which admit a solution
\begin{equation}
\Phi_3=\frac{1}{2}\ln(C'-32g_1r),\qquad A=-2\ln(C'-32g_1r)
\end{equation}
with an integration constant $C'$. This solution gives a domain wall
metric
\begin{equation}
 ds^2=(C'-32g_1r)^{-4}dx^2_{1,1}+dr^2\, .
\end{equation}
\indent For $\Phi_2=0$ which satisfies equation \eqref{Eq2}, a more
general solution can be found by treating $\Phi_1$ as an independent
variable. After combing equations \eqref{Eq3} and \eqref{Eq4} with
\eqref{Eq1}, we can solve for $\Phi_3$ and $A$ as functions of
$\Phi_1$. The result is then substituted in equation \eqref{Eq1} to
find the solution for $\Phi_1(r)$. The resulting solution is given
by
\begin{eqnarray}
\Phi_3&=&\frac{1}{4}\ln\left[\frac{C_1-\coth\Phi_1}{\tanh\Phi_1-C_1}\right],\nonumber \\
A&=&\ln(1-e^{4\Phi_1})-\frac{1}{2}\ln\left[(1+4C_1)^2-(1-4C_1)^2e^{4\Phi_1}\right]\nonumber \\
& &-\frac{1}{2}\ln\left[\left[(1-4C_1)^2e^{4\Phi_1}-(1+4C_1)^2\right]g_1^2-16C_1^2(e^{4\Phi_1}-1)g_2^2\phantom{\sqrt{\sinh(\Phi_1)}}\right.\nonumber \\
&
&\left.-16C_1g_1g_2e^{2\Phi_1}\sqrt{\sinh(2\Phi_1)\left[8C_1\cosh(2\Phi_1)-(1+16C_1^2)\sinh(2\Phi_1)\right]}\right],
\nonumber\\
128C_1g_1^2r&=&g_1\sqrt{8C_1\coth(2\Phi_1)-16C_1^2-1}+2C_1g_2\ln\sinh(2\Phi_1)\nonumber \\
&
&-2C_1g_2\ln\left[(g_1^2+(g_1^2+g_2^2)16C_1^2)\sinh(2\Phi_1)-8C_1g_1^2\cosh(2\Phi_1)\right]
\nonumber \\
&
&-4C_1g_2\tanh^{-1}\left[\frac{g_1\sqrt{8C_1\coth(2\Phi_1)-16C_1^2-1}}{4C_1g_2}\right].
\end{eqnarray}
In the above equations, we have neglected additive integration
constants in $\Phi_1(r)$ and $A(\Phi_1)$ solutions.
\\
\indent Apart from these solutions, we are not able to completely
solve the BPS equations with all scalars non-vanishing in an
analytic form. We will however give a partial result on this
solution since it might be useful for further investigation. It
turns out that combining equations \eqref{Eq1} and \eqref{Eq2} gives
an equation for $\Phi_2$ as a function of $\Phi_1$ with a solution
\begin{equation}
\Phi_2=\frac{1}{4}\ln\left[\frac{1-e^{C}\sinh(2\Phi_1)}{1+e^{C}\sinh(2\Phi_1)}\right].
\end{equation}
If all of the integration constants are set to zero, a simple
solution for $\Phi_3$ can also be found
\begin{equation}
\Phi_3=\cosh^{-1}\left[\frac{1}{2}\sqrt{2-\textrm{csch}(2\Phi_1)\sqrt{\cosh(4\Phi_1)-3}}\right].
\end{equation}
With these relations, equation \eqref{Eq4} would in principle give a
solution for $A(\Phi_1)$ while equation \eqref{Eq1} would give a
solution for $\Phi_1(r)$. We have not succeeded in obtaining an
analytic form for these solutions.

\section{Conclusions}\label{conclusions}
In this paper, we have constructed $N=4$ gauged supergravity in
three dimensions with $SO(4)\ltimes \mathbf{T}^6$ gauge group and
$E_{6(2)}/SU(6)\times SU(2)$ scalar manifold. We have studied some
of the maximally supersymmetric and half-supersymmetric vacua of
this gauged supergravity. The $N=4$ $AdS_3$ critical point with
$SO(4)$ symmetry corresponds to an $N=(4,0)$ SCFT in two dimensions.
We have given the full scalar mass spectrum at this critical point
which might be useful for other holographic applications. In
addition, some half-supersymmetric domain wall solutions have also
been explicitly given in an analytic form. According to the DW/QFT
correspondence \cite{DW/QFT_townsend}, we expect these solutions to
be dual to $N=(4,0)$ non-conformal field theories in two dimensions.
\\
\indent We have also identified two flat directions of the scalar
potential corresponding to exactly marginal deformations of the
$N=(4,0)$ SCFT that preserve all supersymmetries and $SO(4)$
symmetry. Remarkably, these flat directions are not Goldstone
bosons. This is in contrast to the four-dimensional $N=4$ gauged
supergravity studied in \cite{Jan_N4_4D}. In that case, all flat
directions correspond to Goldstone bosons. It would be interesting
to identify the $N=(4,0)$ SCFT and two-dimensional non-conformal
field theories dual to the vacua identified here. Further
investigations of the scalar potential in other scalar sectors
invariant under smaller residual gauge symmetry could be useful for
the study of other deformations of the dual $N=(4,0)$ SCFT in
particular relevant deformations given by scalars in
$(\mathbf{1},\mathbf{3})$, $(\mathbf{2},\mathbf{2})$ and
$(\mathbf{3},\mathbf{3})$ representations.
\\
\indent Due to the equivalence between the gauged supergravity
constructed here and the Yang-Mills gauged supergravity with $SO(4)$
gauge group, it is possible that this theory might be obtained from
higher dimensions. The ungauged $N=4$ supergravity with
$E_{6(2)}/SU(6)\times SU(2)$ scalar manifold can be obtained from a
reduction on a 3-torus ($T^3$) of the minimal supergravity in six
dimensions coupled to three tensor and four vector multiplets. The
three tensor multiplets consist of three scalars parametrized by the
$SO(3,1)/SO(3)$ coset manifold. After dimensional reduction and
dualization of the vector fields coming from the six-dimensional
metric and the (anti) self-dual tensor fields, the resulting
three-dimensional supergravity consists of $40$ scalars as required
by the dimension of $E_{6(2)}/SU(6)\times SU(2)$ coset manifold.
\\
\indent We then expect that the $SO(4)\ltimes \mathbf{T}^6$ gauged supergravity
constructed here should arise from a dimensional reduction of this
six-dimensional supergravity on a 3-shpere ($S^3$). Along the line of this uplifting, it could be useful to compute vector and fermion masses and match with the $AdS_3\times S^3$ spectrum of $N=(1,0)$ six-dimensional supergravity carried out in \cite{de Boer_AdS3_S3}. It should also
be remarked here that, when coupled to hypermultiplets with
hyper-scalars described by $\mc{M}_-$ manifold, the six-dimensional
supergravity could give rise to three-dimensional gauged
supergravity with two scalar target manifolds $\mc{M}_+\times
\mc{M_-}$. It would be interesting to construct an explicit
reduction ansatz similar to the recent result in
\cite{Henning_S3_to_3D}. The result will be very useful in uplifting
the three-dimensional solutions to higher dimensions. We leave this
issue and related ones for future works.
\acknowledgments This work is supported by Chulalongkorn University
through Ratchadapisek Sompoch Endowment Fund under grant
GF-58-08-23-01 (Sci-Super II).
\appendix

\section{Generators of $E_{6(2)}$ and relevant subgroups}\label{E6_generator}
$E_6$ generators in the fundamental representation, used throughout
this paper, have been constructed in \cite{F4} and \cite{E6}. All of
these $78$ generators are denoted by $c_i$, $i=1,\ldots 78$ and are
normalized according to
\begin{equation}
\textrm{Tr}(c_ic_j)=-6\delta_{ij}\, .
\end{equation}
In order to construct the non-compact form $E_{6(2)}$ from the
compact $E_6$, we first identify the maximal compact subgroup $H=
SU(6)\times SU(2)$ with the $SU(2)$ factor corresponding to the
$SU(2)_+$ subgroup of the full $SO(4)$ R-symmetry. We then apply the
``Weyl unitarity trick'' to the remaining $40$ generators to make
them non-compact.
\subsection{$SU(6)\times SU(2)$ subgroup and non-compact generators}
The R-Symmetry group $SU(2)_+\sim SO(3)_+$ is generated by
\begin{equation}
T^{12}=-\frac{1}{2}(c_{51}+c_{78}),\qquad
T^{13}=\frac{1}{2}(c_{52}-c_{77}),\qquad
T^{23}=\frac{1}{2}(c_{36}+\tilde{c}_{53}).
\end{equation}
\indent The generators of the group $H'=SU(6)$ are given by
\begin{eqnarray}
X_i&=&c_i,\qquad i=1,\ldots , 15, \nonumber \\
X_{16}&=&\frac{1}{\sqrt{2}}(c_{52}+c_{77}),\qquad
X_{17}=\frac{1}{\sqrt{2}}(c_{51}-c_{78}),\qquad
X_{18}=\frac{1}{\sqrt{2}}(\tilde{c}_{53}-c_{36}),\nonumber \\
X_{19}&=&\frac{1}{\sqrt{2}}(c_{22}+c_{60}),\qquad
X_{20}=\frac{1}{\sqrt{2}}(c_{23}-c_{59}),\qquad
X_{21}=\frac{1}{\sqrt{2}}(c_{24}+c_{61}),\nonumber \\
X_{22}&=&\frac{1}{\sqrt{2}}(c_{25}-c_{58}),\qquad
X_{23}=\frac{1}{\sqrt{2}}(c_{26}+c_{57}),\qquad
X_{24}=\frac{1}{\sqrt{2}}(c_{27}+c_{55}),\nonumber \\
X_{25}&=&\frac{1}{\sqrt{2}}(c_{28}-c_{54}),\qquad
X_{26}=\frac{1}{\sqrt{2}}(c_{29}-c_{56}),\qquad
X_{27}=\frac{1}{\sqrt{2}}(c_{37}+c_{64}),\nonumber \\
X_{28}&=&\frac{1}{\sqrt{2}}(c_{38}-c_{66}),\qquad
X_{29}=\frac{1}{\sqrt{2}}(c_{39}-c_{62}),\qquad
X_{30}=\frac{1}{\sqrt{2}}(c_{40}+c_{67}),\nonumber \\
X_{31}&=&\frac{1}{\sqrt{2}}(c_{41}+c_{63}),\qquad
X_{32}=\frac{1}{\sqrt{2}}(c_{42}-c_{65}),\qquad
X_{33}=\frac{1}{\sqrt{2}}(c_{43}-c_{69}),\nonumber \\
X_{34}&=&\frac{1}{\sqrt{2}}(c_{44}+c_{68}),\qquad
X_{35}=\tilde{c}_{70}
\end{eqnarray}
where $\tilde{c}_{53}$ and $\tilde{c}_{70}$ generators are defined
by
\begin{equation}
\tilde{c}_{53}=\frac{1}{2}c_{53}+\frac{\sqrt{3}}{2}c_{70},\qquad
\tilde{c}_{70}=-\frac{\sqrt{3}}{2}c_{53}+\frac{1}{2}c_{70}\, .
\end{equation}
It is useful to note that the $SU(4)\times SU(2)\times U(1)\sim
SO(6)\times SO(3)\times U(1)$ subgroup of $SU(6)$ is generated by
$X_i$, $i=1,\ldots, 15$, $(X_{16}, X_{17}, X_{18})$ and $X_{35}$,
respectively.
\\
\indent With the compact $H$ generators defined above, the $40$
non-compact generators are accordingly given by
\begin{eqnarray}
Y_{1}&=&\frac{i}{2}(c_{22}-c_{60}),\qquad
Y_{2}=\frac{i}{2}(c_{23}+c_{59}),\qquad
Y_{3}=\frac{i}{2}(c_{24}-c_{61}),\nonumber \\
Y_{4}&=&\frac{i}{2}(c_{25}+c_{58}),\qquad
Y_{5}=\frac{i}{2}(c_{26}-c_{57}),\qquad
Y_{6}=\frac{i}{2}(c_{27}-c_{55}),\nonumber \\
Y_{7}&=&\frac{i}{2}(c_{28}+c_{54}),\qquad
Y_{8}=\frac{i}{2}(c_{29}+c_{56}),\qquad
Y_{9}=\frac{i}{2}(c_{37}-c_{64}),\nonumber \\
Y_{10}&=&\frac{i}{2}(c_{38}+c_{66}),\qquad
Y_{11}=\frac{i}{2}(c_{39}+c_{62}),\qquad
Y_{12}=\frac{i}{2}(c_{40}-c_{67}),\nonumber \\
Y_{13}&=&\frac{i}{2}(c_{41}-c_{63}),\qquad
Y_{14}=\frac{i}{2}(c_{42}+c_{65}),\qquad
Y_{15}=\frac{i}{2}(c_{43}+c_{69}),\nonumber \\
Y_{16}&=&\frac{i}{2}(c_{44}-c_{68}),\qquad
Y_{17}=\frac{i}{2}(c_{16}+c_{45}),\qquad
Y_{18}=-\frac{i}{2}(c_{17}+c_{46}),\nonumber \\
Y_{19}&=&\frac{i}{2}(c_{18}+c_{47}),\qquad
Y_{20}=-\frac{i}{2}(c_{19}+c_{48}),\qquad
Y_{21}=-\frac{i}{2}(c_{20}+c_{49}),\nonumber \\
Y_{22}&=&-\frac{i}{2}(c_{21}+c_{50}),\qquad
Y_{23}=\frac{i}{2}(c_{30}-c_{71}),\qquad
Y_{24}=\frac{i}{2}(c_{72}-c_{31}),\nonumber \\
Y_{25}&=&\frac{i}{2}(c_{32}-c_{73}),\qquad
Y_{26}=\frac{i}{2}(c_{74}-c_{33}),\qquad
Y_{27}=\frac{i}{2}(c_{75}-c_{34}),\nonumber \\
Y_{28}&=&\frac{i}{2}(c_{76}-c_{35}),\qquad
Y_{29}=\frac{i}{2}(c_{16}-c_{45}),\qquad
Y_{30}=\frac{i}{2}(c_{46}-c_{17}),\nonumber \\
Y_{31}&=&\frac{i}{2}(c_{18}-c_{47}),\qquad
Y_{32}=\frac{i}{2}(c_{48}-c_{19}),\qquad
Y_{33}=\frac{i}{2}(c_{49}-c_{20}),\nonumber \\
Y_{34}&=&\frac{i}{2}(c_{50}-c_{21}),\qquad
Y_{35}=\frac{i}{2}(c_{30}+c_{71}),\qquad
Y_{36}=-\frac{i}{2}(c_{31}+c_{72}),\nonumber \\
Y_{37}&=&\frac{i}{2}(c_{32}+c_{73}),\qquad
Y_{38}=-\frac{i}{2}(c_{33}+c_{74}),\qquad
Y_{39}=-\frac{i}{2}(c_{34}+c_{75}),\nonumber \\
Y_{40}&=&-\frac{i}{2}(c_{35}+c_{76})\, .
\end{eqnarray}
\indent For advantages of future investigations, we give the
non-compact generators corresponding to scalar fields that are
singlets under various subgroups of the $SO(4)$ gauge symmetry. The
following results can be obtained by further decompositions of the
$SO(4)$ representations given in \eqref{SO4_rep}.
\begin{itemize}
\item $SO(3)_{\textrm{diag}}\subset SO(3)\times SO(3)\sim SO(4)$ singlets:
\begin{eqnarray}
\hat{Y}_1&=&Y_{17}+Y_{19}+Y_{24}+Y_{29}-Y_{31}+Y_{36},\nonumber \\
\hat{Y}_2&=&Y_4+Y_7+Y_{12}-Y_{15},\qquad \hat{Y}_3=Y_{26}-Y_{38},\nonumber \\
\hat{Y}_4&=&Y_6-Y_8+Y_{14}+Y_{16},\qquad \hat{Y}_5=Y_{27}-Y_{39},\nonumber \\
\hat{Y}_6&=&Y_2+Y_3+Y_{10}-Y_{11},\qquad \hat{Y}_7=Y_{28}-Y_{40},\nonumber \\
\hat{Y}_8&=&Y_1+Y_5-Y_{9}+Y_{13}
\end{eqnarray}
\item $SU(2)\times SO(2)_s$ singlets:
\begin{eqnarray}
\hat{Y}_1&=&Y_{17}+Y_{19}+Y_{24}+Y_{26}+Y_{29}-Y_{31}+Y_{36}-Y_{38},\nonumber \\
\hat{Y}_2&=&Y_{18}+Y_{20}-Y_{23}-Y_{25}+Y_{30}-Y_{32}-Y_{35}+Y_{37},\nonumber \\
\hat{Y}_3&=&Y_{1}-Y_{4}+Y_{5}-Y_{7},\qquad
\hat{Y}_4=Y_{2}+Y_{3}-Y_{6}+Y_{8},\nonumber \\
\hat{Y}_5&=&Y_{9}+Y_{12}-Y_{13}-Y_{15},\qquad
\hat{Y}_6=Y_{10}-Y_{11}-Y_{14}-Y_{16}
\end{eqnarray}
\item $SU(2)_s$ singlets:
\begin{eqnarray}
\hat{Y}_1&=&Y_{1}-Y_{4}+Y_{5}-Y_{7}-Y_{9}-Y_{12}+Y_{13}+Y_{15},\nonumber \\
\hat{Y}_2&=&Y_2+Y_3-Y_{6}+Y_{8}+Y_{10}-Y_{11}-Y_{14}-Y_{16},\nonumber \\
\hat{Y}_3&=&Y_{17}+Y_{24}-Y_{31}-Y_{38},\qquad
\hat{Y}_4=Y_{18}-Y_{23}+Y_{32}-Y_{37},\nonumber \\
\hat{Y}_5&=&Y_{19}+Y_{26}+Y_{29}+Y_{36},\qquad
\hat{Y}_6=Y_{20}-Y_{25}-Y_{30}+Y_{35}
\end{eqnarray}
\end{itemize}
The $SU(2)_s$ denotes the $SU(2)$ subgroup of $SO(4)$ generated by
self-dual $SO(4)$ generators with $SO(2)_s\subset SU(2)_s$.
\subsection{$SO(6,4)\times U(1)$ subgroup}
The $U(1)$ is generated by $\hat{X}=\tilde{c}_{70}$ while the
$SO(6,4)$ generators are given by
\begin{eqnarray}
\tilde{X}^{1 2}&=& c_1,\qquad \tilde{X}^{1 3}= -c_2,\qquad
\tilde{X}^{2 3}= c_3,\qquad \tilde{X}^{3 4}=
c_6,\nonumber \\
\tilde{X}^{1 4}&=& c_4,\qquad \tilde{X}^{2 4}= -c_5,\qquad
\tilde{X}^{1 5}= c_7,\qquad \tilde{X}^{2 5}= -c_8,\nonumber \\
\tilde{X}^{3 5}&=& c_9,\qquad \tilde{X}^{4 5}= -c_{10},\qquad
\tilde{X}^{5 6}= -c_{15},\qquad \tilde{X}^{1 6}= c_{11},
\nonumber \\
\tilde{X}^{2  6}&=& -c_{12},\qquad \tilde{X}^{4 6}= -c_{14},\qquad
\tilde{X}^{3 6}= c_{13},\qquad \tilde{X}^{1
7}= ic_{16},\nonumber \\
\tilde{X}^{27}&=& -ic_{17},\qquad \tilde{X}^{4 7}= -ic_{19},\qquad
\tilde{X}^{3 7}= ic_{18},\qquad \tilde{X}^{6 7}= -ic_{21},\nonumber \\
\tilde{X}^{ 5 7}&=& -ic_{20},\qquad \tilde{X}^{7 8}= -c_{36},\qquad
\tilde{X}^{1 8}= ic_{30},\qquad \tilde{X}^{2 8}= -ic_{31},
\nonumber \\
\tilde{X}^{4 8}&=& -ic_{33},\qquad \tilde{X}^{3 8}= ic_{32},\qquad
\tilde{X}^{6 8}=
-ic_{35},\qquad \tilde{X}^{5 8}= -ic_{34},\nonumber \\
\tilde{X}^{2 9}&=& -ic_{46},\qquad \tilde{X}^{1 9}= ic_{45}\qquad
\tilde{X}^{4 9}= -ic_{48},\qquad \tilde{X}^{3 9}= ic_{47},\nonumber \\
\tilde{X}^{6 9}&=& -ic_{50},\qquad \tilde{X}^{5 9}= -ic_{49},\qquad
\tilde{X}^{8 9}= -c_{52},\qquad \tilde{X}^{7 9}=
-c_{51},\nonumber \\
\tilde{X}^{1,10}&=&-ic_{71},\qquad \tilde{X}^{2,10}=ic_{72},\qquad \tilde{X}^{3,10}=-ic_{73},\qquad \tilde{X}^{4,10}=ic_{74},\nonumber \\
\tilde{X}^{5,10}&=&ic_{75},\qquad
\tilde{X}^{6,10}=ic_{76},\qquad \tilde{X}^{7,10}=c_{77},\qquad \tilde{X}^{8,10}=c_{78},\nonumber \\
\tilde{X}^{9,10}&=&-\tilde{c}_{53}\, .
\end{eqnarray}
All generators are labelled by $SO(6,4)$ adjoint indices with
$\tilde{X}^{\hat{i}\hat{j}}=-\tilde{X}^{\hat{j}\hat{i}}$,
$\hat{i},\hat{j}=1,\ldots, 10$. The compact subgroup $SO(6)\times
SO(4)$ is generated by $\tilde{X}^{\hat{i}\hat{j}}$,
$\hat{i},\hat{j}=1,\ldots, 6$ and $\tilde{X}^{\hat{i}\hat{j}}$,
$\hat{i},\hat{j}=7,\ldots, 10$, respectively. This coincides with
the $SO(6)\times SO(3)\subset SU(6)$ together with the $SO(3)$
R-symmetry. The $24$ non-compact generators are identified with
$\tilde{X}^{\hat{i},\hat{j}+6}$ for $\hat{i}=1,\ldots, 6$ and
$\hat{j}=1,\ldots, 4$.

\section{Essential formulae}\label{detail}
For a general symmetric space of the form $G/H$ with $G$ and
$H=SO(N)\times H'$ being the global and local symmetry groups, the
$G$ algebra is given by
\begin{eqnarray}
\left[T^{IJ},T^{KL}\right]&=&-4\delta^{[I[K}T^{L]J]}, \qquad
\left[T^{IJ},Y^A\right]=-\frac{1}{2}f^{IJ,AB}Y_B, \nonumber \\
\left[X^\alpha,X^\beta\right]&=&f^{\alpha
\beta}_{\phantom{as}\gamma}X^\gamma,\qquad
\left[X^\alpha,Y^A\right]=h^{\alpha
\phantom{a}A}_{\phantom{a}B}Y^B, \nonumber \\
\left[Y^{A},Y^{B}\right]&=&\frac{1}{4}f^{AB}_{IJ}T^{IJ}+\frac{1}{8}C_{\alpha\beta}h^{\beta
AB}X^\alpha\, . \label{Galgebra}
\end{eqnarray}
$C_{\alpha\beta}$ is an invariant tensor of $H'$, and
$h^{\alpha}_{AB}$ are antisymmetric tensors that commute with
$f^{IJ}_{AB}$.
\\
\indent Using the above algebra, we find that components of $f^{IJ}$
tensor written in flat coset space indices are given by
\begin{equation}
f^{IJ}_{AB}=-\frac{2}{3}\textrm{Tr}(Y^B\left[T^{IJ},Y^A\right]).
\end{equation}
In term of the coset representative, various components of the
$\mc{V}$ map can be computed by using the relations
\begin{eqnarray}
\mc{{V_{\mathbf{a}}}}^{ab,IJ}&=&-\frac{1}{3}\textrm{Tr}(L^{-1}J^{ab}LT^{IJ}),\qquad
\mc{{V_{\mathbf{b}}}}^{ab,IJ}
=-\frac{1}{3}\textrm{Tr}(L^{-1}t^{ab}LT^{IJ}),\nonumber
\\
\mc{{V_{\mathbf{a}}}}^{ab,A}&=&\frac{1}{3}\textrm{Tr}(L^{-1}J^{ab}LY^{A}),
\qquad
\mc{{V_{\mathbf{b}}}}^{ab,A}=\frac{1}{3}\textrm{Tr}(L^{-1}t^{ab}LY^{A}).
\end{eqnarray}
The T-tensors are then obtained from
\begin{eqnarray}
T^{IJ,KL}&=&g_1\left(\mc{{V_{\mathbf{a}}}}^{ab,IJ}\mc{{V_{\mathbf{b}}}}^{cd,KL}+
\mc{{V_{\mathbf{b}}}}^{ab,IJ}\mc{{V_{\mathbf{a}}}}^{cd,KL}\right)\epsilon_{abcd}\nonumber
\\
& &+
g_2\mc{{V_{\mathbf{b}}}}^{ab,IJ}\mc{{V_{\mathbf{b}}}}^{cd,KL}\epsilon_{abcd},\nonumber
\\
T^{IJ,A}&=&g_1\left(\mc{{V_{\mathbf{a}}}}^{ab,IJ}\mc{{V_{\mathbf{b}}}}^{cd,A}+
\mc{{V_{\mathbf{b}}}}^{ab,IJ}\mc{{V_{\mathbf{a}}}}^{cd,A}\right)\epsilon_{abcd}\nonumber
\\
& &+
g_2\mc{{V_{\mathbf{b}}}}^{ab,IJ}\mc{{V_{\mathbf{b}}}}^{cd,A}\epsilon_{abcd}\,
.
\end{eqnarray}
From these relations, the tensors $A_1^{IJ}$, $A^{IJ}_{2i}$ and the
scalar potential can be straightforwardly computed.

\section{Field equations for $SO(4)$ singlet scalars and the metric}\label{field_eq}
In this appendix, we explicitly verify that the BPS equations given in \eqref{Eq1}, \eqref{Eq2}, \eqref{Eq3}, and \eqref{Eq4} indeed satisfy the corresponding second-order field equations. With only scalar fields and the metric, the Lagrangian of the gauged supergravity read, in our convention,
\begin{equation}
\mc{L}=\sqrt{-g}\left[\frac{1}{2}R-\frac{1}{2}P^A_\mu P^{A\mu}-V\right]
\end{equation}
where the scalar potential is given in \eqref{singlet_potential}. The scalar kinetic term is written in term of the coset vielbein $e^A_i$ as
\begin{equation}
P^A_\mu=\pd_\mu \phi^i e^A_i\, .
\end{equation}
It should be noted that, with the relation $g_{ij}=e^A_ie^A_j$, the scalar kinetic term is the same as that given in \cite{dewit}
\begin{equation}
-\frac{1}{2}P^A_\mu P^{A\mu}=-\frac{1}{2}g_{ij}\pd_\mu \phi^i \pd^\mu\phi^j\, .
\end{equation}
In the present case, the coset vielbein can be computed from \eqref{cosetFormula1}
\begin{equation}
P^A_\mu=\frac{1}{3}\textrm{Tr}(L^{-1}\pd_\mu LY^A)\, .
\end{equation}
For completeness, we will explicitly give the result here
\begin{equation}
-\frac{1}{2}P^A_\mu P^{A\mu}=-4\Phi_3'^2-4\cosh^2(2\Phi_3)\left[\cosh^2(2\Phi_2)\Phi_1'^2+\Phi_2'^2\right].\label{scalar_kin}
\end{equation}
\indent
Einstein's equation coming from the above Lagrangian is given by
\begin{equation}
R_{\mu\nu}-\frac{1}{2}g_{\mu\nu}R=P^A_\mu P^A_\nu-g_{\mu\nu}\left[\frac{1}{2}P^A_\rho P^{A\rho}+V\right]\,
\end{equation}
For the metric ansatz \eqref{metric_ansatz}, non-vanishing components of the Ricci tensor and Ricci scalar are the following
\begin{eqnarray}
R_{\mu\nu}&=&-e^{2A}\eta_{\mu\nu}(A''+2A'^2),\nonumber \\
R_{rr}&=&-2(A''+A'^2),\nonumber \\
R&=&-4A''-6A'^2
\end{eqnarray}
for $\mu,\nu=0,1$. These together with the scalar potential \eqref{singlet_potential} and equation \eqref{scalar_kin} imply that all components of the Einstein's equation are satisfied.
\\
\indent For scalar field equations, it is more convenient to write the scalar Lagrangian as
\begin{equation}
\mc{L}_{\textrm{scalar}}=-e^{2A}\left[\frac{1}{2}P^A_rP^{Ar}+V\right].
\end{equation}
From this, the scalar field equations are given by
\begin{equation}
\frac{d}{dr}\frac{\pd \mc{L}_{\textrm{scalar}}}{\pd \Phi'_i}-\frac{\pd \mc{L}_{\textrm{scalar}}}{\pd \Phi_i}=0,\qquad i=1,2,3\, .
\end{equation}
The resulting equations are quite complicated, so we refrain from giving their explicit form here. It can however be verified that all of these equations are satisfied by the BPS equations \eqref{Eq1}, \eqref{Eq2}, \eqref{Eq3}, and \eqref{Eq4}.


\end{document}